# Longitudinal laboratory testing tied to PCR diagnostics in COVID-19 patients reveals temporal evolution of distinctive coagulopathy signatures

Colin Pawlowski[1], Tyler Wagner[1], Arjun Puranik[1], Karthik Murugadoss[1], Liam Loscalzo[1], AJ Venkatakrishnan[1], Rajiv K. Pruthi[2], Damon E. Houghton[2], John C. O'Horo[2], William G. Morice II[2], John Halamka[2], Andrew D. Badley[2], Elliot S. Barnathan[3], Hideo Makimura[3], Najat Khan[3], Venky Soundararajan[1*]

1. nference, inc., One Main Street, Suite 400, East Arcade, Cambridge, MA 02142, USA
2. Mayo Clinic, Rochester MN, USA
3. Janssen pharmaceutical companies of Johnson & Johnson (J&J), Spring House, PA, USA

* Address correspondence to VS (venky@nference.net)

**Temporal inference from laboratory testing results and their triangulation with clinical outcomes as described in the associated unstructured text from the providers' notes in the Electronic Health Record (EHR) is integral to advancing precision medicine. Here, we studied 181 COVID$_{pos}$ and 7,775 COVID$_{neg}$ patients subjected to 1.3 million laboratory tests across 194 assays during a two-month observation period centered around their SARS-CoV-2 PCR testing dates. We found that compared to COVID$_{neg}$ at the time of clinical presentation and diagnostic testing, COVID$_{pos}$ patients tended to have higher plasma fibrinogen levels and similarly low platelet counts, with approximately 25% of patients in both cohorts showing outright thrombocytopenia. However, these measures show opposite longitudinal trends as the infection evolves, with declining fibrinogen and increasing platelet counts to levels that are lower and higher compared to the COVID$_{neg}$ cohort, respectively. Our EHR augmented curation efforts suggest a minority of patients develop thromboembolic events after the PCR testing date, including rare cases with disseminated intravascular coagulopathy (DIC), with most patients lacking the platelet reductions typically observed in consumptive coagulopathies. These temporal trends present, for the first time, fine-grained resolution of COVID-19 associated coagulopathy (CAC), via a digital framework that synthesizes longitudinal lab measurements with structured medication data and neural network-powered extraction of outcomes from the unstructured EHR. This study demonstrates how a precision medicine platform can help contextualize each patient's specific coagulation profile over time, towards the goal of informing better personalization of thromboprophylaxis regimen.**

## Introduction

There is a growing body of evidence that severe COVID-19 outcomes may be associated with dysregulated coagulation[1], including stroke, pulmonary embolism, myocardial infarction, and other venous or arterial thromboembolic complications[2]. This so-called COVID-19 associated coagulopathy (CAC) shares similarities with disseminated intravascular coagulation (DIC) and thrombotic microangiopathy but also has distinctive features[3]. Given the significance of CAC to COVID-19 mortality, there is an urgent need for fine-grained resolution into the temporal manifestation of CAC, particularly in comparison to the broad-spectrum of other, better characterized coagulopathies. While there is a growing body of evidence suggesting associations between COVID-19 infection and mortality with thrombocytopenia, D-dimer levels, and prolongation of prothrombin time, the potentially heterogeneous signatures of CAC onset and progression as well as their connection to clinical outcomes are not well defined [1,4,5]. An advanced understanding of this phenotype may aid in risk stratifying patients, thus facilitating optimal monitoring strategies during disease evolution through the paradigm of precision medicine, which hopefully will lead to further improvements in treatment decisions and overall survival.

To this end, we instituted a holistic data science platform across an academic health care system that enables machine intelligence to augment the curation of phenotypes and outcomes from 15.2 million EHR clinical notes and associated 3 million lab tests from 1,192 COVID-19-positive ($COVID_{pos}$) and 47,344 confirmed COVID-19-negative ($COVID_{neg}$) patients over a retrospectively defined 2-month observation period straddling the date of the PCR test (see ***Methods***). By compiling all available laboratory testing data for the 30 days preceding the SARS-CoV-2 PCR diagnostic testing date ('day 0'), as well as the 30 days following the diagnostic testing date, and triangulating this information with medications and clinical outcomes, we were able to identify laboratory abnormalities significantly associated with positive COVID-19 diagnosis. We identified coagulation-related parameters among this set of abnormalities and then studied aggregate as well as individual patient trajectories that could aid in extracting a temporal signature of CAC onset and progression. We also correlated these signals with the clinical outcomes of these patients.

In order to hone into longitudinal lab test trends that would apply at the individual patient level, we restricted our analysis to patients with available serial testing data, which had at least 3 test results of the same type during the observation period. After applying these inclusion criteria, 181 $COVID_{pos}$ and 7,775 $COVID_{neg}$ patients met study inclusion criteria. The median age in the $COVID_{pos}$ and $COVID_{neg}$ groups were 59.9 years and 63.9 years respectively (see ***Methods*** and Table 1), and the numbers of males were 96 (53%) and 4,064 (52%), respectively. The number of pre-existing coagulopathies in the $COVID_{pos}$ and $COVID_{neg}$ groups were 56 (33%) and 3,325 (40%), respectively. Overall, more patients in the $COVID_{neg}$ cohort were hospitalized at the start of the study period. In particular, the number of $COVID_{pos}$ patients hospitalized on the SARS-CoV-2 PCR testing date was 23 (12.7%), compared to 1435 (18.5%) for the $COVID_{neg}$ cohort. In order to balance some of these clinical covariates and others between the two cohorts, we applied 1:10 propensity score matching to define a subset of 1,810 patients from the $COVID_{neg}$ cohort to use for the final statistical analyses (see ***Methods***). In particular, the general categories of covariates considered for balancing included: demographics, anticoagulant/antiplatelet medication use, and hospital admission status. Population-level characteristics of the $COVID_{pos}$, $COVID_{neg}$, and the final propensity score-matched $COVID_{neg}$ cohorts are summarized in **Table 1**.

# Results

### *Longitudinal analysis highlights classes of lab abnormalities in COVID-19 progression*

To identify laboratory test results which differ between $COVID_{pos}$ and $COVID_{neg}$ patients, we analyzed longitudinal trends for 194 laboratory test results in the 30 days prior to and after the day of PCR testing (designated as day 0). As most patients did not undergo laboratory testing for each assay on a daily basis, we grouped the measurements into time windows reflecting potential stages of infection as follows: pre-infection (days -30 to -11), pre-PCR (days -10 to -2), time of clinical presentation (days -1 to 0), and post-PCR phases 1 (days 1 to 3), 2 (days 4-6), 3 (days 7-9), 4 (days 10-12), 5 (days 13-15), and 6 (days 16-30). We only considered test-time window pairs in which there were at least three patients contributing laboratory test results in both groups. During each time window, we then compared the distribution of results from $COVID_{pos}$ versus $COVID_{neg}$ patients, allowing us to identify any lab tests which were significantly altered in $COVID_{pos}$ patients during any time of disease acquisition, onset, and/or progression.

Out of 863 lab test-time window pairs with adequate data points for comparison, we identified 88 such pairs (comprising 36 unique lab tests) which met our thresholds for statistical significance (Cohen's D > 0.35, BH-adjusted Mann-Whitney p-value < 0.05) (**Table 2**). Among these were lab tests which may be considered positive controls for our analysis, including a reduction in blood oxygenation in $COVID_{pos}$ patients after diagnosis and elevated titers of SARS-CoV-2 IgG antibodies in the late phase 6 (days 16-30) after diagnosis (**Figures 1A-B**). We also identified abnormalities in several other classes of lab tests, including immune cell counts (**Figures 1C-G**), erythrocyte-related tests (**Figures 1H-I**), serum chemistry (**Figures 1J-K**), and coagulation-related tests (**Figure 2**).

With respect to coagulation, we found that plasma fibrinogen was significantly elevated in $COVID_{pos}$ patients at the time of diagnosis (Cohen's D = 1.092, BH-adjusted Mann-Whitney p-value = 0.016, **Table 2, Figure 2A**). This hyperfibrinogenemia generally resolved during the 10 days following diagnosis, even declining to levels significantly lower than the $COVID_{neg}$ cohort (**Figure 2A**). Conversely, platelet counts were lower in the $COVID_{pos}$ cohort at the time of clinical presentation but tended to increase over the subsequent 10 days to levels significantly higher than those in $COVID_{neg}$ patients (Cohen's D = 0.361, BH-adjusted Mann-Whitney p-value = 0.008, **Table 2, Figure 2B**). While thrombocytopenia has been reported in COVID-19 patients before[6,7], an upward trend in platelet counts after diagnosis has not been described to our knowledge. Among other coagulation-related tests, we found no difference in prothrombin time (PT) and slight elevations in activated partial thromboplastin times (aPTT) after diagnosis (**Figure 2C**). D-dimer levels were frequently above normal limits in both the $COVID_{pos}$ and $COVID_{neg}$ cohorts and were not significantly different between these cohorts during any time windows (**Figure 2D**). Of note, the observed trends in both fibrinogen and platelet counts could not be attributed to the administration of either antiplatelet or anticoagulant therapies, as drug class-matched cohorts of $COVID_{neg}$ patients failed to show these trends after the day of PCR testing (**Figures 2E-F**).

We also performed similar analyses comparing the $COVID_{pos}$ and $COVID_{neg}$ cohorts using different time windows definitions including daily trends (**Figure 3**). This approach offers the advantage of increased granularity at the cost of sample size per time point, but we did identify similar lab tests as altered in $COVID_{pos}$ patients using each approach including the fibrinogen decline and platelet increase in the $COVID_{pos}$ cohort after diagnosis (**Figure 3**).

### ***Thrombosis is an enriched phenotype among COVID19 patients undergoing longitudinal lab testing***

Given the recently described coagulopathies associated with COVID-19[1,2,3], we were intrigued by these temporal trends in fibrinogen levels and platelet counts in the $COVID_{pos}$ cohort. We next asked whether the observed coagulation-related laboratory trends were associated with clinical manifestations of thrombosis. To do so, we employed a BERT-based neural network[8] (see **Methods**) to identify patients who experienced a thrombotic event after their SARS-CoV-2 testing date. Specifically, we extracted diagnostic sentiment from EHR notes (e.g. whether a patient was diagnosed with a phenotype, suspected of having a phenotype, ruled out for having a phenotype, or other) regarding specific thromboembolic phenotypes including pulmonary embolism (PE), deep vein thrombosis (DVT), venous thromboembolism (VTE), thrombotic stroke, myocardial infarction, and cerebral venous thrombosis.

We found that 40 of the total 1729 $COVID_{pos}$ cohort (2.3%) were positively diagnosed with one or more of these phenotypes in the 30 days after PCR testing, with the majority of these patients (27 of 40; 68%) experiencing a DVT. Interestingly, we found that after creating subsets of the patients with longitudinal lab testing data (i.e. the patients meeting the criteria for inclusion in our study), 33 of the 181 patients (18%) had at least one EHR-derived clot diagnosis, including 25 patients with DVT (**Table 6**). Thus, the cohort under consideration here is highly enriched (**Table 4**; hypergeometric p-value < $1x10^{-20}$) for patients experiencing thrombotic events compared to the overall $COVID_{pos}$ cohort.

***Longitudinal platelet trends are not strongly associated with the development of thrombosis in COVID-19 patients***

Among the 181 $COVID_{pos}$ patients with longitudinal lab testing data, only 10 were serially tested starting at clinical presentation for fibrinogen versus 87 tested for platelets. As such, we first analyzed whether associations exist between platelet counts (or temporal alterations thereof) and clotting propensity in this cohort. Among these 87, there were 68 patients without thrombosis after PCR-based diagnosis ("non-thrombotic") and 19 patients with thrombosis ("thrombotic"). Thrombocytopenia (platelet count $<150x10^9/L$) was observed in ~25% of $COVID_{pos}$ (and $COVID_{neg}$) patients at the time of diagnosis (**Figure 4A**), but the platelet levels at this time point were not associated with the subsequent formation of a blood clot in the $COVID_{pos}$ cohort (**Figure 4B**).

We hypothesized that the previously discussed increase in platelet counts after COVID-19 diagnosis may be associated with the development of blood clots. If true, then we would expect the thrombotic $COVID_{pos}$ cohort to show significantly higher maximum platelet counts during their course of disease progression compared to the non-thrombotic $COVID_{pos}$ cohort. We found that this was not the case, as maximum platelet counts were similar in the two groups (**Figure 4C**). Similarly, among the 74 $COVID_{pos}$ patients with at least one post-diagnosis platelet count higher than that at the time of presentation, the degree of maximal platelet increase was not associated with the development of thrombosis (**Figure 4D**). It would certainly be of interest to perform this same analysis on the larger $COVID_{pos}$ cohort (n = 1729; 40 thrombotic vs. 1689 non-thrombotic), but we were not able to do so given the lack of longitudinal testing available for the large majority of non-thrombotic $COVID_{pos}$ patients (**Table 3**).

Conversely, we explored whether some $COVID_{pos}$ patients may experience clotting in the setting of low or declining platelets (e.g., consumptive coagulopathy) despite the population-level trend of increasing platelets over time. Indeed, we found that 4 of 19 thrombotic patients showed absolute platelet counts below $100x10^9/L$ during at least one post-diagnosis time window (below dotted red line in **Figure 4E**). We next identified all $COVID_{pos}$ patients who showed at least one post-diagnosis platelet count lower than the count at the time of diagnosis. Interestingly, the binary presence of post-diagnosis platelet reductions was associated with thrombotic status (n = 27 of 68 in non-thrombotic population; n = 14 of 19 in thrombotic population; p = 0.0017). While the maximum degree of absolute platelet reduction was not associated with clot development overall (p = 0.15; **Figure 4F**), we did find that 3 of the 15 thrombotic patients experienced a reduction of at least $100x10^9$/L relative to the time of diagnosis (below gray line in **Figure 4F**). Of note, similar fractions of non-thrombotic $COVID_{pos}$ patients also showed these low or declining platelet counts, indicating that these trends are not specific indicators of thrombosis (**Figures 4E-F**).

***Consumptive coagulopathy contributes to a small fraction of COVID-19 associated thromboses***

The observed declining platelet counts and thrombocytopenia in the context of thrombosis in a small fraction of $COVID_{pos}$ patients is consistent with previous reports that fewer than 1% of survivors, but over 70% of non-survivors, meet the International Society on Thrombosis and Hemostasis (ISTH) criteria for disseminated intravascular coagulation (DIC)[1]. As was previously noted, hyperfibrinogenemia was among the strongest lab test features distinguishing $COVID_{pos}$ from $COVID_{neg}$ patients at diagnosis, but the subsequent downward trend (**Figure 2B**) could be attributed to a resolving acute phase response and/or consumption of fibrinogen in a systemic coagulopathy. Using our BERT-based sentiment extraction, we found that only 2 of the 1729 $COVID_{pos}$ patients were diagnosed with DIC, all of whom were included in our longitudinal cohort of 181 $COVID_{pos}$ patients. Both extracted diagnoses were manually confirmed by EHR review. This finding suggests that declining fibrinogen after COVID-19 diagnosis typically represents a physiologic return to normal range rather than pathologic coagulation factor consumption. To further examine the plasma fibrinogen trends among COVID-19 patients with DIC, with non-DIC thrombosis, and without thrombosis, we examined patient-level lab test trends from the 10 individuals who were tested for fibrinogen both at the time of diagnosis and at least two times subsequently.

This patient-level analysis indeed revealed multiple distinct trajectories with respect to fibrinogen and other coagulation parameters in $COVID_{pos}$ patients. Four of these ten individuals developed at least one blood clot during their hospital course. Only one was identified by our BERT model (and confirmed by manual EHR review) to have low-grade DIC, and as expected we found this patient's longitudinal lab test pattern to be consistent with consumptive coagulopathy (Patient 124; **Figure 5A**). At the time of diagnosis, this patient showed significant hyperfibrinogenemia with elevated D-dimers (1304.5 ng/mL) and a borderline normal platelet count ($153x10^9$/L). Over the next ten days, this patient's fibrinogen levels consistently decreased, reaching a minimum of 110 mg/dL on day 9. Similarly, after an initial recovery to $190x10^9$/L the platelet counts consistently declined starting on day 2 post-diagnosis, reaching a minimum of $117x10^9$/L on day 11. D-dimer levels exponentially increased after 5 days, reaching a maximum of 41,300 ng/mL on day 10. Phenotypically, this patient experienced both thrombotic (right internal jugular vein and right superior thyroid artery) and hemorrhagic (oropharyngeal and pulmonary) events. This combination of lab results and clinical manifestations is consistent with the diagnosis of DIC-like consumptive coagulopathy during the first week after COVID-19 diagnosis.

Lab test results from the other three non-DIC thrombotic patients with longitudinal fibrinogen testing confirm the presence of alternative forms of coagulopathy in the COVID-19 population. Patient 23 developed a clot on day 4 post-diagnosis in the context of a declining fibrinogen level and increasing D-dimers but steady platelet counts, which actually increased shortly thereafter (**Figure 5B**). Patient 79 developed several clots after day 3 post-diagnosis in the setting of upward trending platelets (which eventually exceed the upper limit of normal) and elevated levels of both fibrinogen and D-dimers (**Figure 5C**). Patient 94 developed a clot on day 8 post-diagnosis with relatively stable platelet counts within normal limits and steadily declining fibrinogen levels (**Figure 5D**).

Whether early elevations in plasma fibrinogen contribute to the clotting observed in the non-DIC like $COVID_{pos}$ cohort is unfortunately not possible for us to assess here as only one of the ten patients who received fibrinogen testing at the time of clinical presentation and twice subsequently developed thrombosis, and this patient was diagnosed with DIC. This hypothesis may warrant further analysis in cohorts with more longitudinal fibrinogen data, but again it is important to note that several $COVID_{pos}$ patients who presented with hyperfibrinogenemia did not go on to develop thromboses (**Figures 5E-F**). This emphasizes that a steady post-diagnosis decline in plasma fibrinogen may represent physiologic resolution of acute phase response rather than a pathologic consumption of fibrinogen and other coagulation factors (**Figures 5B, D-F**).

Taken together, this analysis affirms that a DIC-like coagulopathy resulting in a combination of hemorrhage and thrombosis can develop in the setting of COVID-19 infection. However, the observations that (1) DIC was formally diagnosed in only 2 of 1729 $COVID_{pos}$ patients and (2) 32 of 54 thrombotic $COVID_{pos}$ patients with serial platelet testing never showed a single platelet count lower than that obtained on day 0 emphasizes that consumptive coagulopathy is an exception rather than the rule as it pertains to thrombotic phenotypes in COVID-19 patients.

## Discussion

A number of studies on clinical characteristics and lab tests are shedding light on the spectrum of hematological parameters associated with COVID-19 patients. In an initial study of 41 patients from Wuhan, the blood counts in $COVID_{pos}$ patients showed leukopenia and lymphopenia, and prothrombin time and D-dimer levels were higher in ICU patients than in non-ICU patients[9]. Another study based on 343 Wuhan $COVID_{pos}$ patients found that a D-dimer level of at least 2.0 µg/ml could predict mortality with a sensitivity of 92.3% and a specificity of 83.3%[10]. An independent study of 43 COVID-19 patients found significant differences between mild and severe cases in plasma interleukin-6 (IL-6), D-dimers, glucose, thrombin time, fibrinogen, and C-reactive protein ($P < 0.05$)[4]. While such studies indeed highlight that hematological and inflammatory abnormalities are prevalent in $COVID_{pos}$, a high-resolution temporal understanding of how these parameters evolve in COVID-19 patients post diagnosis has not been established. Specifically, in the wake of accumulating evidence for hypercoagulability in $COVID_{pos}$ patients, there are important clinical questions emerging regarding the necessity of and guidelines for thromboprophylaxis in patient management.

The ability to derive this longitudinal understanding of COVID-19 progression, including laboratory abnormalities and their associated clinical manifestations, mandates the synthesis of structured and unstructured EHR data (e.g., lab tests and clinical notes) at a large scale. The fact that tens of thousands of patients have undergone SARS-CoV-2 testing at major academic medical centers (AMCs) provides an abundance of potential data to perform this analysis but also poses significant challenges from a practicality standpoint. Manual review and curation of patient trajectories and associated testing results is not practical, nor is it likely to provide comprehensive or even entirely accurate individual patient records. Rather, triangulation across datasets, including lab measurements, clinical notes, and prescription information, using a scalable digitized approach to extract structured data along with sentiment-surrounded clinical phenotypes and outcomes is required to efficiently perform this analysis in a timely fashion.

By developing and deploying such a digitized platform on the entirety of EHR data from a large AMC, we have identified in an unbiased manner, laboratory abnormalities that characterize $COVID_{pos}$ patients from their $COVID_{neg}$ counterparts who also presented with symptoms leading to the performance of a SARS-CoV-2 PCR test. The abnormalities in coagulation-related tests, including fibrinogen and platelets, were intriguing in the context of literature reporting the occurrence of various clotting phenotypes in COVID-19 patients, including DIC-like consumptive coagulopathies along with more isolated clotting events in the lungs, central nervous system, and other tissues[1,2,3]. Our finding that consumptive coagulopathy represents a minority of COVID-19 associated clotting events corroborates and provides context for other studies which have reported overt DIC or DIC-like disease in over 70% of non-survivors but far lower fractions of survivors[1]. A study of 83 COVID-19 patients suggests that when LMWH prophylaxis was initiated on hospital admission, progression to DIC was rare[11].

While our analysis highlights that consumptive coagulopathy should be considered in the minority of $COVID_{pos}$ patients with significant serial reductions in platelet counts, it remains to be seen whether the post-diagnosis platelet increases or early hyperfibrinogenemia which we observed may contribute mechanistically to the clotting in the much larger non-DIC thrombotic COVID-19 population. It is important to note that despite the trend of increasing platelets, the platelet count only extended above the normal range ($>450 \times 10^9$/L) in 24 of 135 $COVID_{pos}$ patients with serial measurements, and the development of such outright thrombocytosis was observed with similar frequencies in the thrombotic and non-thrombotic cohorts. Further, the fact that several patients with elevated fibrinogen (i.e. >400 mg/dL) at presentation did not develop thromboses suggests that early hyperfibrinogenemia is not a singular driver of subsequent clotting events, but a small sample size (n = 10 patients; 9 non-thrombotic vs. 1 thrombotic) limited the power of this analysis.

Despite these uncertainties, this linking of longitudinal trends to patient outcomes provides several useful pieces of clinical information, including but not limited to: (1) Early hyperfibrinogenemia is to be expected in COVID-19 patients and may not be useful in determining an appropriate thromboprophylaxis approach for a given patient. (2) Declining fibrinogen levels shortly after diagnosis are also expected and likely represent the resolution of an acute phase response in most patients rather than a decline secondary to the onset of consumptive coagulation. (3) Borderline or overt thrombocytopenia is common in COVID-19 patients at the time of clinical presentation, and the initial platelet count does not robustly predict patients who are likely to develop thromboses. (4) After diagnosis, COVID-19 patients generally show an upward trend in platelets. Patients whose platelets trend down after diagnosis should be monitored, as platelet reductions after clinical presentation are associated with thromboses and significant reductions may be indicative of ongoing consumptive coagulopathy.

There certainly remains much to understand regarding the molecular pathophysiology underlying COVID-19 associated thrombosis. While elevations in fibrinogen and D-dimer have been reported (and confirmed by our analyses), the circulating levels of tissue factor and other coagulation factors have not been assessed broadly in this population. Interestingly, single cell RNA-sequencing data highlights that the respiratory tract, including the nasal cavity, are strongly enriched for high expression of tissue factor (encoded by the gene *F3*) (**Figure 6A**). Specifically, *F3* is highly expressed in epithelial cells of these tissues - the same cell types which have been found to express *ACE2*, the putative entry receptor for SARS-CoV-2 (**Figures 6B-C**). Type II pneumocytes, which may also express *ACE2* and have been reported as a putative entry site for SARS-CoV-2[12–14], are also strongly enriched among all human cell types showing high *F3* expression (**Figure 6A**). Whether infection-related damage to these epithelial cells may lead to shedding of tissue factor into circulation or its increased exposure circulating coagulation factors locally in respiratory tissues would be interesting to evaluate. Such a mechanism could at least partially contribute to the development of clots systemically and locally in the pulmonary vasculature. This has been confirmed in at least 3 autopsy series that reported both microthrombi and pulmonary emboli in a significant proportion of COVID-19 patients[15].

In addition to our findings regarding abnormal coagulation parameters in COVID-19 patients, we identified several other laboratory features which differed from the $\text{COVID}_{\text{neg}}$ cohort. For example, others have previously reported leukopenia and T cell deficiencies in COVID-19 patients at presentation [9,16]; we identified a leukopenia which existed both prior to and at the time of clinical presentation which recovered during the week following COVID-19 diagnosis (**Figure 1C**). Specifically, we found that monocytes, basophils, neutrophils, and eosinophils followed this temporal trend (**Figures 1D-G**). With respect to erythrocytes, we found that both $\text{COVID}_{\text{pos}}$ and $\text{COVID}_{\text{neg}}$ patients developed a worsening mild anemia after PCR testing, but the trajectory was actually slower for $\text{COVID}_{\text{pos}}$ patients (**Figures 1B, 1H, Table 1**). This progressive anemia was interestingly paralleled by a declining mean corpuscular volume unique to the $\text{COVID}_{\text{pos}}$ cohort (**Figure 1I**). This may represent an inflammation associated anemia and could simply be related to repeated blood testing (i.e., mild blood loss) in these patients. Among serum chemistry tests, we observed distinct patterns showing reduced calcium in $\text{COVID}_{\text{pos}}$ patients at presentation and through the first week of infection versus elevated magnesium between days 4 and 15 (**Figures 1J-K**). While low calcium at the time of presentation has been previously described in COVID-19 patients [17], elevations in serum magnesium levels in the 1-2 weeks after diagnosis have not been reported. Determining whether these serum chemistry alterations are induced by SARS-CoV-2 or have any impact on viral replication or pathogenicity, although preliminary analyses have shown no association between these lab measurements at any time point and patient mortality (not shown).

It is important to note that while we center the study period around the PCR testing date, this date may not correspond to the same disease state of COVID-19 for each individual in the $\text{COVID}_{\text{pos}}$ cohort. Additionally, there are several covariates which may influence these longitudinal trends and should be explored further. For example, we have already considered whether previous or concomitant administration of anticoagulants or antiplatelet agents influences patient lab test results and/or outcomes. In **Figure 2D-E**, we present the longitudinal platelet counts and fibrinogen levels for subsets of the $\text{COVID}_{\text{pos}}$ and $\text{COVID}_{\text{neg}}$ cohort who are on and off these medications. While we do observe elevated levels of platelets generally in the $COVID_{neg}$ cohort who are on antiplatelets, this potential medication effect does not alone account for the differences in platelet levels that we observe in the $COVID_{pos}$ cohort. Similarly, the authors wish to explore whether longitudinal lab measurement trends differ between outpatient, inpatient, and ICU admitted patient cohorts. New data sets can also be utilized; for example, rather than grouping patients by the identified thromboembolic phenotypes extracted from the clinical notes alone, patients could be stratified by those who had imaging studies (Duplex ultrasound, CT scan, etc.) performed, and phenotypes could be directly extracted from these procedural reports. As more data accumulates from $COVID_{pos}$ and $COVID_{neg}$ patients in the coming months, these analyses will be expanded to assess similarities and differences in the temporal trends of laboratory test results among a wider range of patient subgroups relevant for COVID-19 outcomes, such as those who have pre-existing conditions (e.g., diabetes, hypertension, obesity, malignancies) or patients who are on specific medication (e.g., ACE inhibitors, statins, immunosuppressants).

In summary, this work demonstrates significant progress in our goal of enabling scaled and digitized analyses of longitudinal unstructured and structured EHR records to identify variables (e.g., laboratory results) which are associated with relevant clinical phenotypes (e.g., COVID-19 diagnosis and outcomes). In doing so, we identified trends in lab test results which may be relevant to monitor in COVID-19 patients and warrant both clinical and mechanistic follow-up in more targeted and explicitly controlled prospective analyses.

## Figure Legends

**Figure 1. Longitudinal and temporally resolved analysis highlights positive control lab tests elevated in $COVID_{pos}$ patients along with immune, hematologic, and serum chemistry signatures. (A-K)** Longitudinal trends of various lab tests in $COVID_{pos}$ (red) versus $COVID_{neg}$ (blue) patients. For any windows of time during which at least one patient in each cohort had test results, data is shown as mean with standard errors. Values given horizontally along the top of the plot are BH-adjusted Mann-Whitney test p-values. Values in red correspond to lab test-time point pairs which met the cutoffs for statistical significance (adjusted p-value < 0.05 and Cohen's D > 0.35). Values given horizontally along the bottom of the plot are the numbers of patients in the $COVID_{neg}$ and $COVID_{pos}$ cohorts, respectively (i.e. "# $COVID_{neg}$ | # $COVID_{pos}$).

**Figure 2. $COVID_{pos}$ patients show distinctly opposite temporal trends in platelet counts and fibrinogen starting at the time of diagnosis. (A-B)** Longitudinal trends of platelet counts and fibrinogen in $COVID_{pos}$ (red) versus $COVID_{neg}$ (blue) patients. For any time windows during which at least one patient in each cohort had test results, data is shown as mean with standard errors. Values given horizontally along the top of the plot are BH-adjusted Mann-Whitney test p-values. Values in red correspond to lab test-time point pairs which met the cutoffs for statistical significance (adjusted p-value < 0.01 and Cohen's D > 0.35). Values given horizontally along the bottom of the plot are the numbers of patients in the $COVID_{neg}$ and $COVID_{pos}$ cohorts, respectively (i.e. "# $COVID_{neg}$ | # $COVID_{pos}$). **(C)** Longitudinal trends of other coagulation-related tests including prothrombin time (PT), activated partial thromboplastic time (aPTT), and D-dimers. **(D-E)** Trends in platelet counts and fibrinogen levels at daily resolution with $COVID_{pos}$ and $COVID_{neg}$ cohorts

stratified based on administration of anticoagulant or antiplatelet agents. For each cohort, average lab values and standard errors are shown for each day with at least three observations.

**Figure 3. Aggregate results from 9 selected lab tests for $COVID_{pos}$ and $COVID_{neg}$ (Matched) patient cohorts over a 60-day period centered around the PCR diagnostic testing date (day 0).** Lab tests include: **(A)** Platelets; **(B)** Fibrinogen, Plasma; **(C)** Prothrombin Time, Plasma; **(D)** Activated Partial Thromboplastin Time; **(E)** D-Dimer; **(F)** Magnesium, Serum/Plasma; **(G)** Basophils Absolute; **(H)** Neutrophils, Blood; **(I)** Alkaline Phosphatase, Serum. References ranges are shown at the top of each plot. For each cohort, average lab values and standard errors are shown for each day with at least three observations.

**Figure 4. Association between platelet counts and thrombosis in $COVID_{pos}$ and $COVID_{neg}$ cohorts. (A)** Distribution of platelet counts at the time of PCR testing for $COVID_{pos}$ cohort. Vertical dotted gray lines correspond to upper and lower limits of normal platelet counts ($150\text{-}450 \times 10^9$/L). **(B)** Comparison of platelet counts at time of PCR testing between $COVID_{pos}$ patients who did not (red) and did (blue) subsequently develop thromboses. **(C)** Comparison of maximum platelet counts (considering counts at and after clinical presentation for PCR testing) between $COVID_{pos}$ patients who did not and did subsequently develop thromboses. **(D)** Comparison of maximum degree of platelet increase after clinical presentation in $COVID_{pos}$ patients who did not and did subsequently develop thromboses. **(E)** Comparison of minimum platelet counts (considering counts at and after clinical presentation for PCR testing) between $COVID_{pos}$ patients who did not and did subsequently develop thromboses. **(F)** Comparison of maximum degree of platelet decline after clinical presentation in $COVID_{pos}$ patients who did not and did subsequently develop thromboses.

**Figure 5. Longitudinal analyses of platelet counts, plasma fibrinogen, and D-dimer levels in individual patients with or without thrombotic disease. (A)** Patient 124 developed hemorrhagic and thrombotic phenotypes in the context of declining fibrinogen, declining platelets, and increasing D-dimers. This is consistent with a DIC-like coagulopathy. **(B)** Patient 23 developed clots in the setting of declining fibrinogen and elevated D-dimers but stable platelet counts which increased shortly thereafter. **(C)** Patient 94 developed clots while showing increases in platelet counts along with plasma fibrinogen and D-dimers. **(D)** Patient 94 developed clots with relatively stable platelet counts and steadily declining plasma fibrinogen. **(E)** Patient 13 did not develop clots or bleeding despite coordinate decrease in platelet counts and fibrinogen which may be mistaken for a DIC-like coagulopathy. **(F)** Patient 51 did not develop clots despite showing a post-diagnosis decline in plasma fibrinogen similar to several patients in the thrombotic cohort.

**Figure 6. Single cell RNA-sequencing highlights expression of tissue factor (factor 3, *F3*) in respiratory epithelial cells which also express ACE2. (A)** Across >1.9 million single cell samples, cells from the respiratory tract and nasal cavity are strongly enriched for high expression of the tissue factor (F3 gene). Further, type II pneumocytes are among the cell types that are most strongly enriched for high expression of the tissue factor gene. **(B-C)** Violin plot representations showing expression of tissue factor (B) and ACE2 (C), the putative SARS-CoV-2 entry receptor, in epithelial cells from the respiratory tract and nasal cavity. Data is from a previously published study by Durante, et al.[18]

## Acknowledgments

The authors thank Mathai Mammen, Jim List, JoAnne Foody, Patrick Lenehan, Murali Aravamudan, Rakesh Barve, Sankar Ardhanari, and Vishy Thiagarajan, for their helpful feedback.

## Methods

***Study design, setting and patient population***

This is a retrospective study of patients who underwent polymerase chain reaction (PCR) testing for suspected SARS-CoV-2 infection at the Mayo Clinic and hospitals affiliated to the Mayo health system.

This research was conducted under IRB 20-003278, “Study of COVID-19 patient characteristics with augmented curation of Electronic Health Records (EHR) to inform strategic and operational decisions”. All analysis of EHRs was performed in the privacy-preserving environment secured and controlled by the Mayo Clinic. Nference, the Mayo Clinic, and the Janssen pharmaceutical companies of Johnson & Johnson (J&J) subscribe to the basic ethical principles underlying the conduct of research involving human subjects as set forth in the Belmont Report and strictly ensure compliance with the Common Rule in the Code of Federal Regulations (45 CFR 46) on the Protection of Human Subjects.

***Longitudinal lab testing tied to COVID-19 PCR diagnostic testing***

We analyzed data from 48,536 patients who received PCR tests from the Mayo Clinic between February 15, 2020 to May 8, 2020. Among this population, 1,192 patients had at least one positive SARS-CoV-2 PCR test result, and 47,344 patients had all negative PCR test results. In order to align the data for the analysis of aggregate longitudinal trends, we selected a reference date for each patient. For patients in the $COVID_{pos}$ cohort, we used the date of the first positive PCR test result as the reference date (day = 0). For patients with all negative PCR tests, we used the date of the first PCR test result as the reference date (day = 0). We defined the study period for each patient to be 30 days before and after the PCR testing date. Patients with contradictory PCR test results were excluded for the purpose of this analysis; e.g. a positive PCR test result and a negative PCR test result on the same day, or a positive PCR test result followed immediately by several negative PCR test results.

Over 3 million test results from 5,536 different types of lab tests were recorded for the patients who received PCR tests in the 60-day windows surrounding their PCR testing dates at the Mayo Clinic campuses in Minnesota, Arizona and Florida. Among these lab tests, we restricted our analysis to 194 tests with at least 1000 observations total and at least 10 observations from the $COVID_{pos}$ cohort. In addition, we considered different subsets of the $COVID_{pos}$ cohort for the analysis of each of the 194 lab tests, due to differences in availability of testing results. For each lab test, we consider the results from patients with 3 or more observations during the study period.

In the end, there are 181 COVID-19 positive and 7,775 COVID-19 negative patients that had 3 or more test results during the study period for at least one of the assays among the 194 lab tests considered. We take this set of 181 COVID-19 positive patients to be the $COVID_{pos}$ cohort. In order to construct the $COVID_{neg}$ cohort from the 7,775 COVID-19 negative patients, we apply propensity score matching, which is described in the next section.

### *Propensity score matching to select the final $COVID_{neg}$ cohort*

To construct a $COVID_{neg}$ cohort similar in baseline clinical covariates to the $COVID_{pos}$ cohort, we employ 1:10 propensity score matching[19]. In particular, first we trained a regularized logistic regression model to predict the likelihood that each patient will have a positive or negative COVID-19 test result, using the following covariates: demographics (age, gender, race), anticoagulant/antiplatelet medication use (orders for alteplase, antithrombin III, apixaban, argatroban, aspirin, bivalirudin, clopidogrel, dabigatran, dalteparin, enoxaparin, eptifibatide, heparin, rivaroxaban, warfarin in the past year and in the past 30 days), and hospitalization status (i.e. whether or not the patient was hospitalized on the date of PCR testing).

Using the predictions from the logistic regression model as propensity scores, we then matched each of the 181 patients in the $COVID_{pos}$ cohort to 10 patients out of the 7,775 COVID-19 negative patients, using greedy nearest-neighbor matching without replacement[20].[20] As a result, we ended up with a final $COVID_{neg}$ cohort that included 1,810 patients with similar baseline characteristics to the $COVID_{pos}$ cohort. The characteristics of the two cohorts are summarized in Table S1.

Further, for the analyses conducted on individual lab tests, which include only a subset of patients from the $COVID_{pos}$ cohort, we use the propensity scores to match each patient from the $COVID_{pos}$ cohort to 10 patients from the $COVID_{neg}$ cohort which have the most similar propensity scores and lab tests available. For example, for the Fibrinogen lab test, in which we have data on 10 patients from the $COVID_{pos}$ cohort, we select 100 patients from the $COVID_{neg}$ cohort with Fibrinogen lab tests available and the most similar propensity scores to be the control group. In this way, we ensure that all of the comparisons are done between subsets of the positive and negative cohorts with similar propensity scores, and therefore similar underlying characteristics.

### *Statistical significance assessments over 1-month post SARS-CoV-2 PCR diagnosis date*

To identify tests that show significant differentiation, we first compared the overall distributions of test scores of $COVID_{pos}$ vs $COVID_{neg}$ over the 30-day period following testing.

For each test the procedure was to:

(1) Compute a single score per patient by averaging all their tests over the duration (may be 1 or many tests per patient); we now have a set of patient test scores for the $COVID_{pos}$ patient sets, and one for the $COVID_{neg}$ patient set.

(2) Perform a two-sample *t*-test as well as a Mann-Whitney *U* test to get p-values against the null hypothesis that both sets have the same mean (for the *t*-test), and that both sets come from the same distribution (for the Mann-Whitney *U* test)

(3) Compute Cohen's *d* (effect size) as a measure of the effect size of the difference; for the Mann-Whitney *U* test, the *U* statistic itself serves as an effect size.

We then filtered by *t*-test p-value then ranked by Cohen's *d* to identify which tests appeared to have the most differentiation between $COVID_{pos}$ and $COVID_{neg}$ scores; we then investigated some of these tests in deeper detail by looking at how they changed over time. We applied the same procedure to compute statistics for each of our time windows (restricting tests to the respective duration in each case).

### *Augmented curation of anticoagulant administration and the coagulopathy outcomes from the unstructured clinical notes and their triangulation to structured EHR databases*

A state-of-the-art BERT-based neural network[8] was previously developed to classify sentiment regarding a diagnosis in the EHR[21]. Sentences containing phenotypes were classified into the following categories: Yes (confirmed diagnosis), No (ruled out diagnosis), Maybe (possibility of disease), and Other (alternate context, e.g. family history of disease). The neural network used to perform this classification was trained using nearly 250 different phenotypes and 18,500 sentences and achieves 93.6% overall accuracy and over 95% precision and recall for Yes/No sentiment classification[21]. Here, this model was used to classify the sentiment around coagulopathies in the unstructured text of the 181 $COVID_{pos}$ and 7,775 $COVID_{neg}$ patients' clinical notes, structuring this information so that it could be compiled with longitudinal lab measurement and medication information.

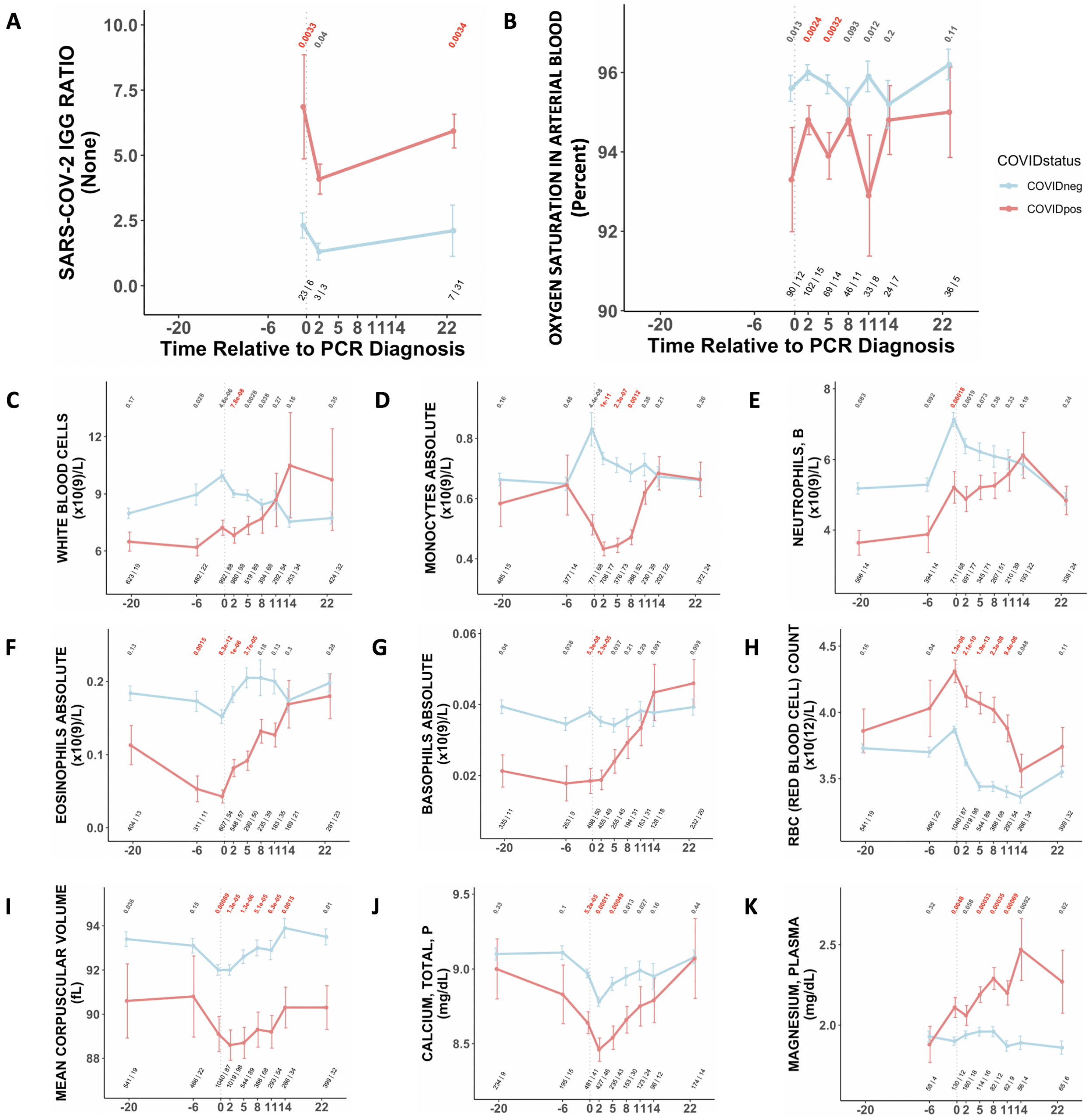
A
SARS-COV-2 IGG RATIO (None)
Time Relative to PCR Diagnosis
B
OXYGEN SATURATION IN ARTERIAL BLOOD (Percent)
Time Relative to PCR Diagnosis
COVIDstatus
COVIDneg
COVIDpos
C
WHITE BLOOD CELLS (x10(9)/L)
D
MONOCYTES ABSOLUTE (x10(9)/L)
E
NEUTROPHILS, B (x10(9)/L)
F
EOSINOPHILS ABSOLUTE (x10(9)/L)
G
BASOPHILS ABSOLUTE (x10(9)/L)
H
RBC (RED BLOOD CELL) COUNT (x10(12)/L)
I
MEAN CORPUSCULAR VOLUME (fL)
J
CALCIUM, TOTAL, P (mg/dL)
K
MAGNESIUM, PLASMA (mg/dL)

**Figure 1**

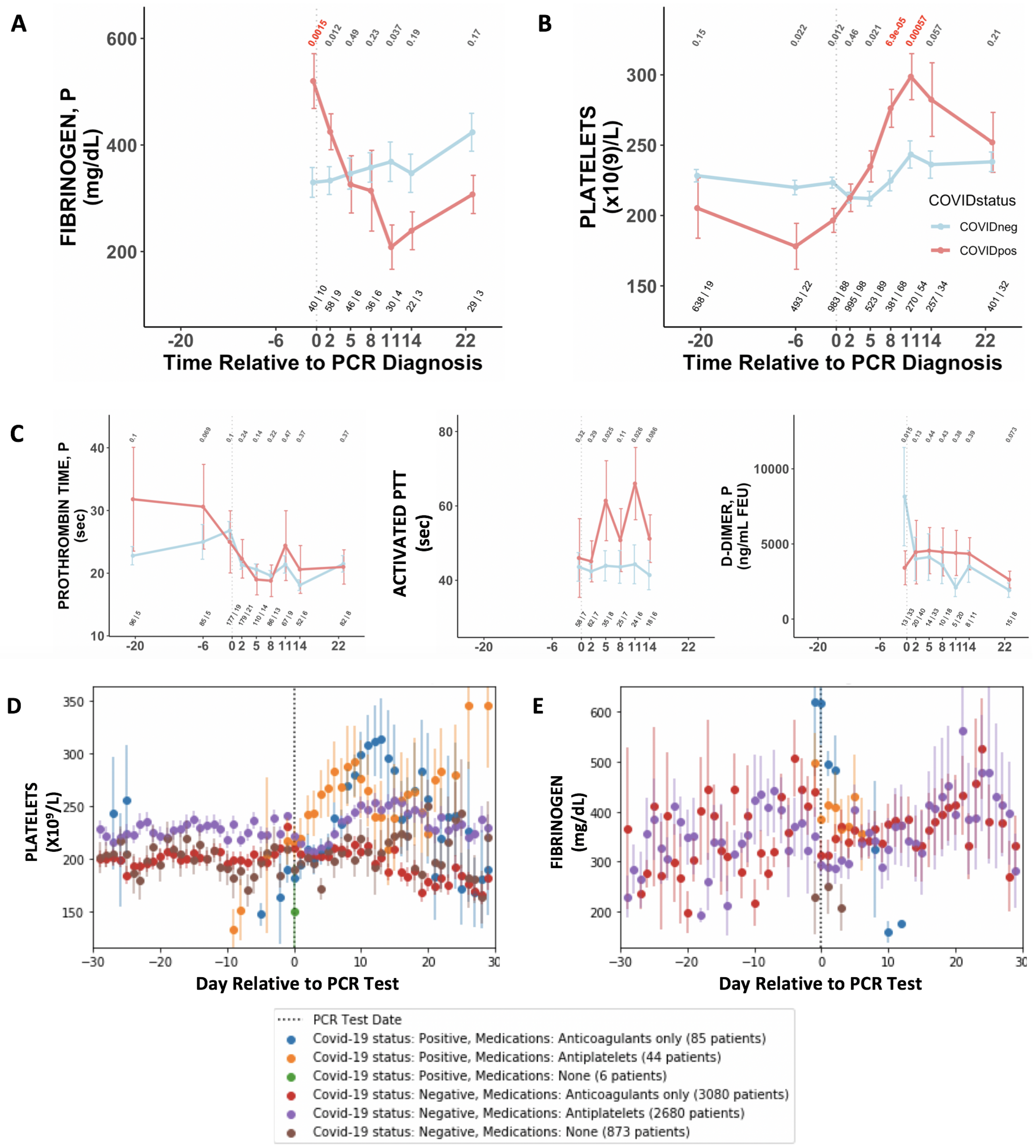
A
FIBRINOGEN, P (mg/dL)
Time Relative to PCR Diagnosis
B
PLATELETS (x10(9)/L)
COVIDstatus
COVIDneg
COVIDpos
Time Relative to PCR Diagnosis
C
PROTHROMBIN TIME, P (sec)
ACTIVATED PTT (sec)
D-DIMER, P (ng/mL FEU)
D
PLATELETS (X10⁹/L)
Day Relative to PCR Test
E
FIBRINOGEN (mg/dL)
Day Relative to PCR Test
PCR Test Date
Covid-19 status: Positive, Medications: Anticoagulants only (85 patients)
Covid-19 status: Positive, Medications: Antiplatelets (44 patients)
Covid-19 status: Positive, Medications: None (6 patients)
Covid-19 status: Negative, Medications: Anticoagulants only (3080 patients)
Covid-19 status: Negative, Medications: Antiplatelets (2680 patients)
Covid-19 status: Negative, Medications: None (873 patients)

**Figure 2**

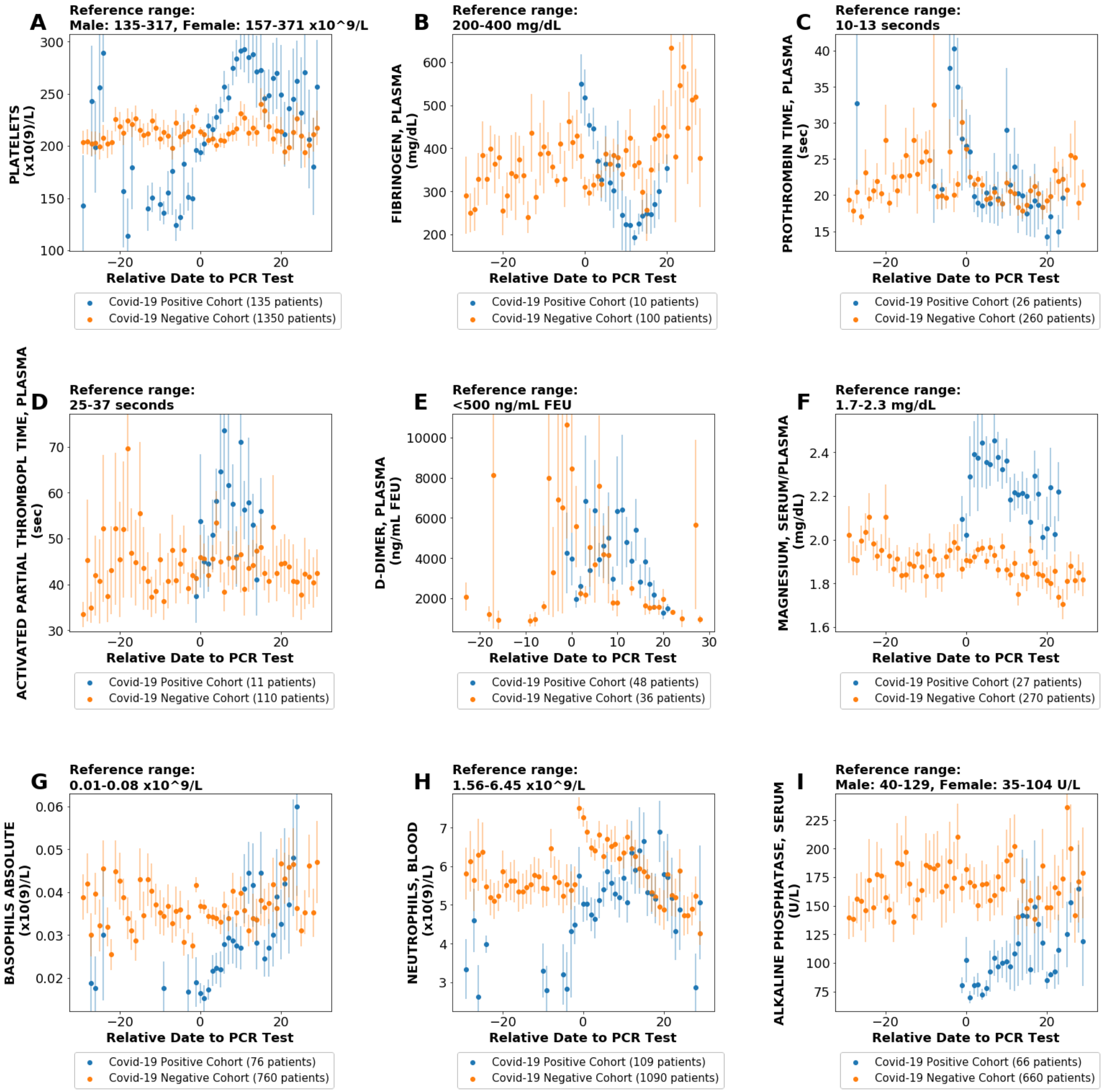

A
Reference range:
Male: 135-317, Female: 157-371 x10^9/L
PLATELETS (x10(9)/L)
Relative Date to PCR Test
Covid-19 Positive Cohort (135 patients)
Covid-19 Negative Cohort (1350 patients)
B
Reference range:
200-400 mg/dL
FIBRINOGEN, PLASMA (mg/dL)
Relative Date to PCR Test
Covid-19 Positive Cohort (10 patients)
Covid-19 Negative Cohort (100 patients)
C
Reference range:
10-13 seconds
PROTHROMBIN TIME, PLASMA (sec)
Relative Date to PCR Test
Covid-19 Positive Cohort (26 patients)
Covid-19 Negative Cohort (260 patients)
D
Reference range:
25-37 seconds
ACTIVATED PARTIAL THROMBOPL TIME, PLASMA (sec)
Relative Date to PCR Test
Covid-19 Positive Cohort (11 patients)
Covid-19 Negative Cohort (110 patients)
E
Reference range:
<500 ng/mL FEU
D-DIMER, PLASMA (ng/mL FEU)
Relative Date to PCR Test
Covid-19 Positive Cohort (48 patients)
Covid-19 Negative Cohort (36 patients)
F
Reference range:
1.7-2.3 mg/dL
MAGNESIUM, SERUM/PLASMA (mg/dL)
Relative Date to PCR Test
Covid-19 Positive Cohort (27 patients)
Covid-19 Negative Cohort (270 patients)
G
Reference range:
0.01-0.08 x10^9/L
BASOPHILS ABSOLUTE (x10(9)/L)
Relative Date to PCR Test
Covid-19 Positive Cohort (76 patients)
Covid-19 Negative Cohort (760 patients)
H
Reference range:
1.56-6.45 x10^9/L
NEUTROPHILS, BLOOD (x10(9)/L)
Relative Date to PCR Test
Covid-19 Positive Cohort (109 patients)
Covid-19 Negative Cohort (1090 patients)
I
Reference range:
Male: 40-129, Female: 35-104 U/L
ALKALINE PHOSPHATASE, SERUM (U/L)
Relative Date to PCR Test
Covid-19 Positive Cohort (66 patients)
Covid-19 Negative Cohort (660 patients)

**Figure 3**

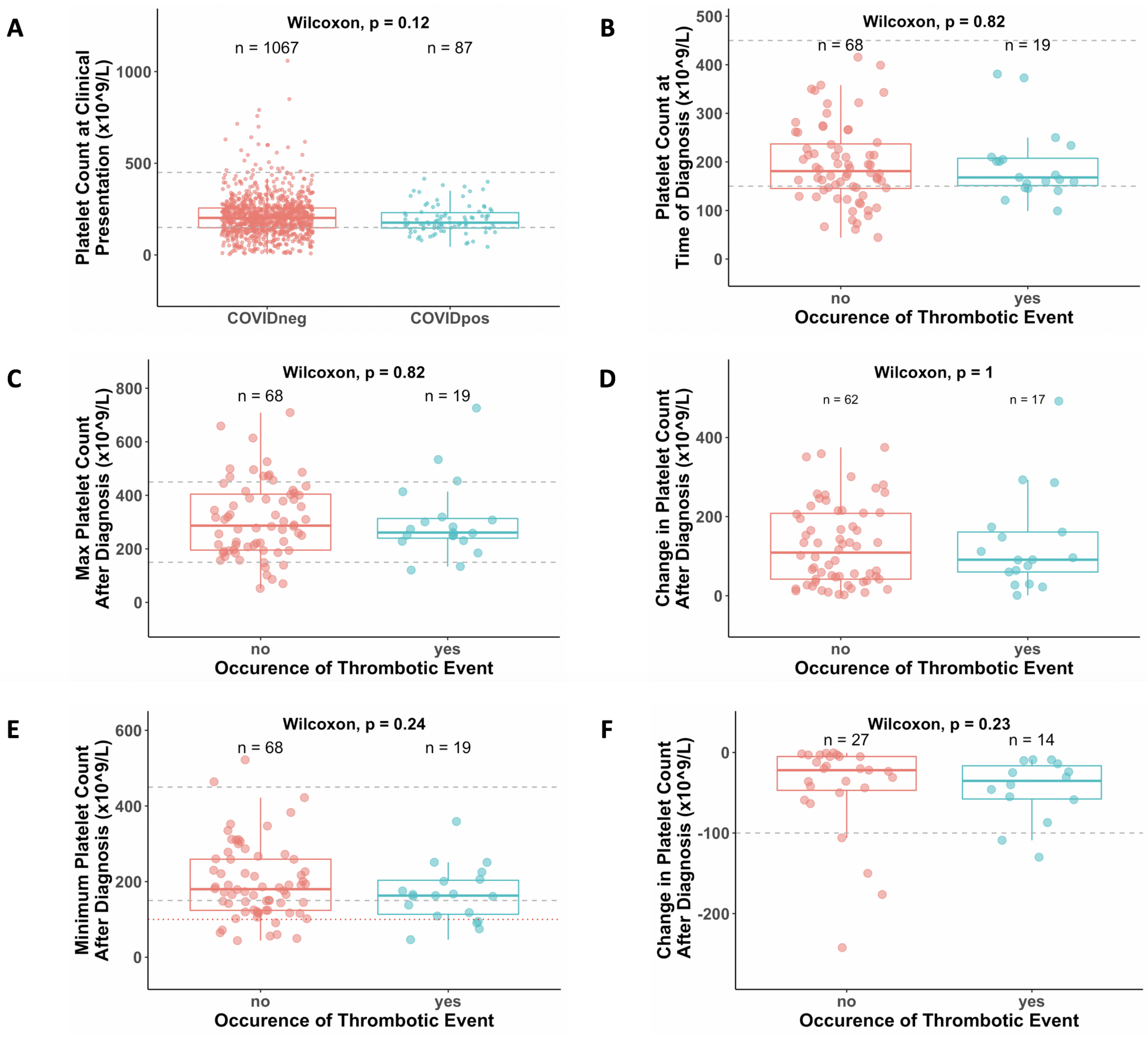
A
Wilcoxon, p = 0.12
n = 1067
n = 87
Platelet Count at Clinical Presentation (x10^9/L)
COVIDneg
COVIDpos
B
Wilcoxon, p = 0.82
n = 68
n = 19
Platelet Count at Time of Diagnosis (x10^9/L)
no
yes
Occurence of Thrombotic Event
C
Wilcoxon, p = 0.82
n = 68
n = 19
Max Platelet Count After Diagnosis (x10^9/L)
no
yes
Occurence of Thrombotic Event
D
Wilcoxon, p = 1
n = 62
n = 17
Change in Platelet Count After Diagnosis (x10^9/L)
no
yes
Occurence of Thrombotic Event
E
Wilcoxon, p = 0.24
n = 68
n = 19
Minimum Platelet Count After Diagnosis (x10^9/L)
no
yes
Occurence of Thrombotic Event
F
Wilcoxon, p = 0.23
n = 27
n = 14
Change in Platelet Count After Diagnosis (x10^9/L)
no
yes
Occurence of Thrombotic Event

**Figure 4**

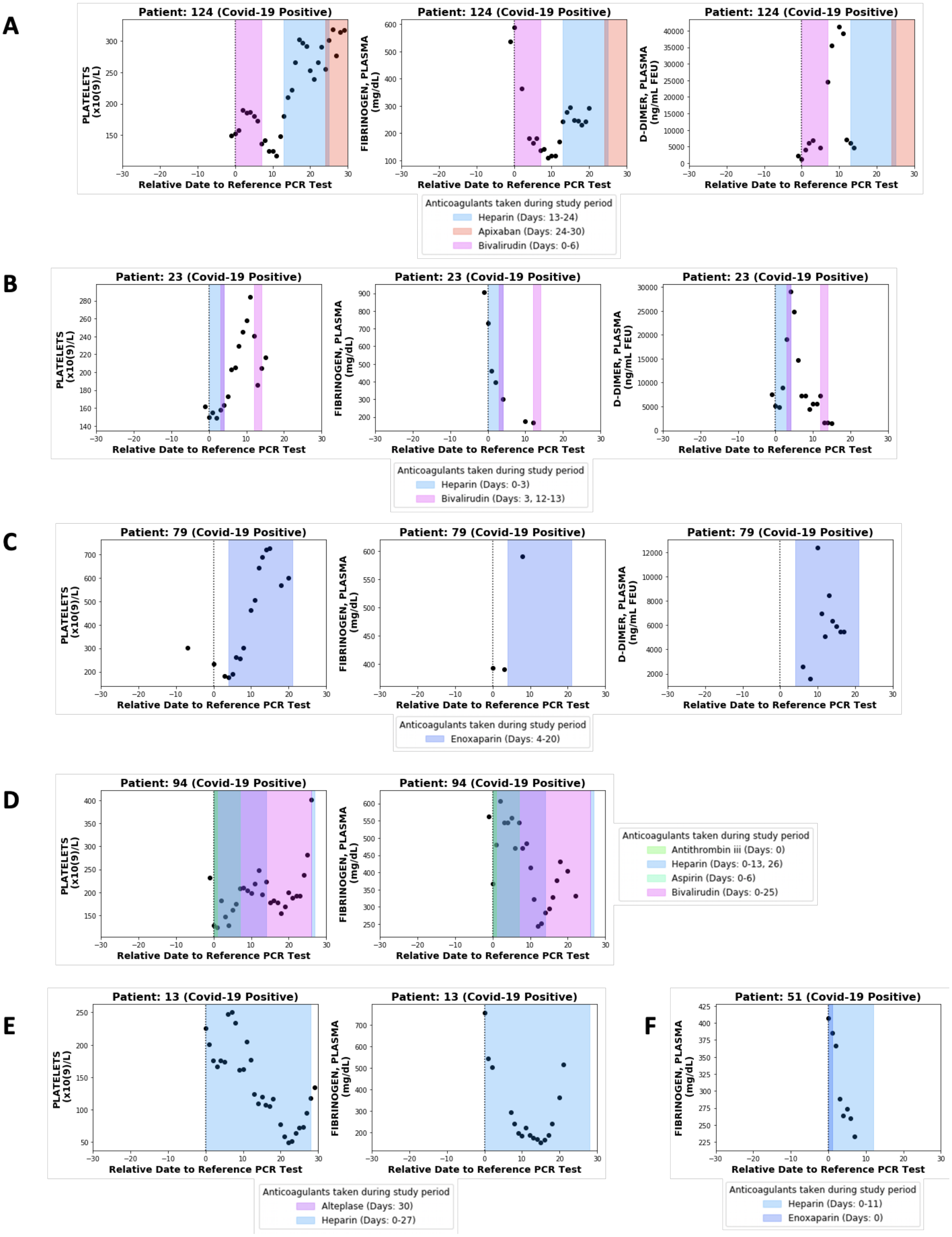

A
Patient: 124 (Covid-19 Positive)
PLATELETS (x10(9)/L)
FIBRINOGEN, PLASMA (mg/dL)
D-DIMER, PLASMA (ng/mL FEU)
Relative Date to Reference PCR Test
Anticoagulants taken during study period
Heparin (Days: 13-24)
Apixaban (Days: 24-30)
Bivalirudin (Days: 0-6)
B
Patient: 23 (Covid-19 Positive)
PLATELETS (x10(9)/L)
FIBRINOGEN, PLASMA (mg/dL)
D-DIMER, PLASMA (ng/mL FEU)
Relative Date to Reference PCR Test
Anticoagulants taken during study period
Heparin (Days: 0-3)
Bivalirudin (Days: 3, 12-13)
C
Patient: 79 (Covid-19 Positive)
PLATELETS (x10(9)/L)
FIBRINOGEN, PLASMA (mg/dL)
D-DIMER, PLASMA (ng/mL FEU)
Relative Date to Reference PCR Test
Anticoagulants taken during study period
Enoxaparin (Days: 4-20)
D
Patient: 94 (Covid-19 Positive)
PLATELETS (x10(9)/L)
FIBRINOGEN, PLASMA (mg/dL)
Relative Date to Reference PCR Test
Anticoagulants taken during study period
Antithrombin iii (Days: 0)
Heparin (Days: 0-13, 26)
Aspirin (Days: 0-6)
Bivalirudin (Days: 0-25)
E
Patient: 13 (Covid-19 Positive)
PLATELETS (x10(9)/L)
FIBRINOGEN, PLASMA (mg/dL)
Relative Date to Reference PCR Test
Anticoagulants taken during study period
Alteplase (Days: 30)
Heparin (Days: 0-27)
F
Patient: 51 (Covid-19 Positive)
FIBRINOGEN, PLASMA (mg/dL)
Relative Date to Reference PCR Test
Anticoagulants taken during study period
Heparin (Days: 0-11)
Enoxaparin (Days: 0)

**Figure 5**

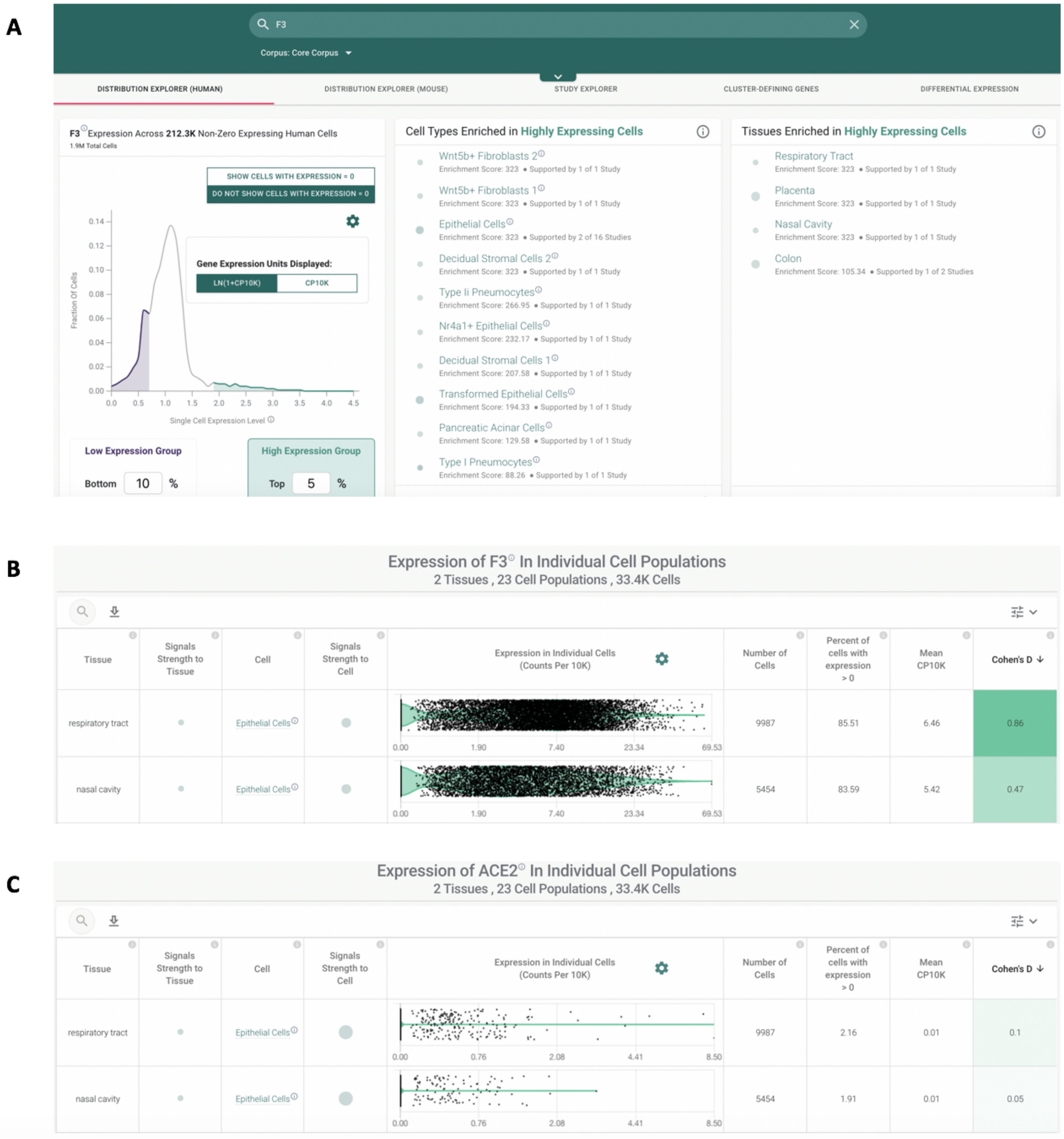
A
B
C
Expression of F3 In Individual Cell Populations
2 Tissues , 23 Cell Populations , 33.4K Cells
Expression of ACE2 In Individual Cell Populations
2 Tissues , 23 Cell Populations , 33.4K Cells

**Figure 6**

**Table 1. Summary of patient characteristics for the overall $COVID_{pos}$, $COVID_{neg}$ (matched), and $COVID_{neg}$ cohorts.** The $COVID_{neg}$ (matched) cohort was constructed using 1:10 propensity score matching to balance each of the clinical covariates, including: demographics (age, gender, race), medication use (anticoagulant/antiplatelet use within 30 days/1 year of PCR testing date), and hospitalization status on the date of PCR testing.

| Patient Characteristics | $COVID_{pos}$ | $COVID_{neg}$ (matched) | $COVID_{neg}$ |
|---|---|---|---|
| ***Number of patients*** | 181 | 1810 | 7775 |
| ***Age in years*** | 59.9 | 59.7 | 63.9 |
| ***Gender*** | | | |
| Male | 96 (53.0%) | 956 (52.8%) | 4064 (52.3%) |
| ***Race:*** | | | |
| White | 129 (71.3%) | 1329 (73.4%) | 6907 (88.8%) |
| Black | 12 (6.6%) | 149 (8.2%) | 362 (4.7%) |
| Asian | 9 (5.0%) | 95 (5.2%) | 140 (1.8%) |
| American Indian | 11 (6.1%) | 48 (2.7%) | 48 (0.6%) |
| Other | 20 (11.0%) | 189 (10.4%) | 318 (4.1%) |
| ***Medication use in the preceding 30 days of PCR testing date:*** | | | |
| Anticoagulants | 29 (16.0%) | 265 (14.6%) | 2466 (31.7%) |
| Antiplatelets | 24 (13.3%) | 249 (13.8%) | 1498 (19.3%) |
| ***Medication use in the preceding 1 year of PCR testing date:*** | | | |
| Anticoagulants | 35 (19.3%) | 377 (20.8%) | 3404 (43.8%) |
| Antiplatelets | 35 (19.3%) | 360 (19.9%) | 2461 (31.7%) |
| ***Hospitalized on PCR testing date*** | 23 (12.7%) | 229 (12.7%) | 1435 (18.5%) |

**Table 2. Summary of lab tests significantly different between $COVID_{pos}$ and propensity score matched $COVID_{neg}$ cohorts during at least one clinical time window.** Data from individual patients was averaged over the defined time windows, and mean values were compared between $COVID_{pos}$ and $COVID_{neg}$ patients. The lab test-time window pairs shown are those with at least three patients contributing data in both cohorts which met our defined thresholds for statistical significance (cohen's D > 0.35 and BH-adjusted Mann Whitney p-value < 0.05). Rows are sorted by the relative time window (from earliest to latest), and then by the adjusted Mann-Whitney p-value. Sample sources are denoted as: P = Plasma, S = Serum, S/P = Serum/Plasma, B = Blood, U = Urine.

| Test | Units | Time Window | Count $COVID_{pos}$ | Count $COVID_{neg}$ | Mean $COVID_{pos}$ | Mean $COVID_{neg}$ | Cohen's D | Mann Whit U | BH-adj MW p-value |
|---|---|---|---|---|---|---|---|---|---|
| EOSINOPHILS ABSOLUTE | x10(9)/L | Pre-diagnosis | 11 | 311 | 0.0531 | 0.173 | -0.502 | 0.24 | 0.016 |
| EOSINOPHILS ABSOLUTE | x10(9)/L | Clinical presentation | 54 | 607 | 0.0427 | 0.152 | -0.499 | 0.22 | 2.8767E-09 |
| CALCIUM, TOTAL, S | mg/dL | Clinical presentation | 46 | 541 | 8.22 | 8.93 | -0.991 | 0.21 | 5.3938E-09 |
| BASOPHILS ABSOLUTE | x10(9)/L | Clinical presentation | 50 | 498 | 0.0185 | 0.0379 | -0.656 | 0.27 | 3.6512E-06 |
| RBC (RED BLOOD CELL) COUNT | x10(12)/L | Clinical presentation | 87 | 1040 | 4.31 | 3.87 | 0.534 | 0.35 | 5.9047E-05 |
| HEMATOCRIT, B | % | Clinical presentation | 87 | 1049 | 38.1 | 34.9 | 0.473 | 0.37 | 0.0005 |
| ALBUMIN, S/P | g/dL | Clinical presentation | 38 | 378 | 3.26 | 3.75 | -0.733 | 0.3 | 0.001 |
| CALCIUM, TOTAL, P | mg/dL | Clinical presentation | 41 | 481 | 8.64 | 8.97 | -0.535 | 0.32 | 0.001 |
| RED CELL DISTRIBUTION WIDTH CV | % | Clinical presentation | 70 | 863 | 14.1 | 15 | -0.427 | 0.36 | 0.001 |
| NEUTROPHILS, B | x10(9)/L | Clinical presentation | 68 | 711 | 5.21 | 7.13 | -0.371 | 0.37 | 0.003 |
| HEMOGLOBIN, B | g/dL | Clinical presentation | 88 | 963 | 12.4 | 11.5 | 0.378 | 0.39 | 0.005 |
| LACTATE DEHYDROGENASE, S | U/L | Clinical presentation | 19 | 81 | 467.2 | 340.5 | 0.456 | 0.25 | 0.006 |
| SODIUM, S | mmol/L | Clinical presentation | 46 | 559 | 137 | 138.7 | -0.437 | 0.36 | 0.009 |
| MEAN CORPUSCULAR VOLUME | fL | Clinical presentation | 87 | 1040 | 89.1 | 92 | -0.386 | 0.4 | 0.011 |
| CARBOXYHEMOGLOBIN, ARTERIAL | % | Clinical presentation | 16 | 137 | 0.478 | 0.93 | -0.705 | 0.27 | 0.016 |
| FIBRINOGEN, P | mg/dL | Clinical presentation | 10 | 40 | 520 | 329.8 | 1.092 | 0.19 | 0.016 |
| TROPONIN T, 5TH GEN, P | ng/L | Clinical presentation | 12 | 27 | 75.7 | 581.2 | -0.58 | 0.21 | 0.019 |
| MAGNESIUM, PLASMA | mg/dL | Clinical presentation | 12 | 130 | 2.11 | 1.9 | 0.704 | 0.27 | 0.041 |
| BICARBONATE, P | mmol/L | Clinical presentation | 41 | 482 | 22.4 | 23.8 | -0.35 | 0.38 | 0.044 |
| EOSINOPHILS PERCENT | % | Clinical presentation | 7 | 63 | 0.3 | 1.4 | -0.799 | 0.21 | 0.05 |
| MONOCYTES ABSOLUTE | x10(9)/L | Days 1-3 Post-Dx | 77 | 708 | 0.433 | 0.733 | -0.581 | 0.27 | 2.8767E-09 |
| RBC (RED BLOOD CELL) COUNT | x10(12)/L | Days 1-3 Post-Dx | 98 | 1019 | 4.12 | 3.62 | 0.678 | 0.31 | 3.6246E-08 |
| HEMATOCRIT, B | % | Days 1-3 Post-Dx | 98 | 1016 | 36.3 | 32.7 | 0.591 | 0.33 | 2.6849E-06 |
| CALCIUM, TOTAL, S | mg/dL | Days 1-3 Post-Dx | 55 | 587 | 8.26 | 8.75 | -0.787 | 0.28 | 3.6512E-06 |
| WHITE BLOOD CELLS | x10(9)/L | Days 1-3 Post-Dx | 98 | 980 | 6.82 | 9.01 | -0.367 | 0.34 | 4.8081E-06 |
| EOSINOPHILS ABSOLUTE | x10(9)/L | Days 1-3 Post-Dx | 57 | 548 | 0.0815 | 0.182 | -0.41 | 0.31 | 5.0765E-05 |
| ALKALINE PHOSPHATASE, S | U/L | Days 1-3 Post-Dx | 40 | 371 | 78.2 | 149.9 | -0.394 | 0.29 | 0.0002 |
| HEMOGLOBIN, B | g/dL | Days 1-3 Post-Dx | 98 | 993 | 11.8 | 10.9 | 0.461 | 0.37 | 0.0003 |
| MEAN CORPUSCULAR VOLUME | fL | Days 1-3 Post-Dx | 98 | 1019 | 88.6 | 92 | -0.459 | 0.37 | 0.0004 |
| BASOPHILS ABSOLUTE | x10(9)/L | Days 1-3 Post-Dx | 49 | 455 | 0.0188 | 0.0352 | -0.57 | 0.32 | 0.001 |
| RED CELL DISTRIBUTION WIDTH CV | % | Days 1-3 Post-Dx | 78 | 829 | 14.2 | 15.1 | -0.407 | 0.36 | 0.001 |
| ALBUMIN, S/P | g/dL | Days 1-3 Post-Dx | 40 | 367 | 3.24 | 3.59 | -0.623 | 0.31 | 0.001 |
| CALCIUM, TOTAL, P | mg/dL | Days 1-3 Post-Dx | 46 | 427 | 8.46 | 8.78 | -0.533 | 0.33 | 0.002 |
| CALCIUM, IONIZED, B | mg/dL | Days 1-3 Post-Dx | 10 | 110 | 4.34 | 4.72 | -0.941 | 0.15 | 0.003 |
| ALKALINE PHOSPHATASE, P | U/L | Days 1-3 Post-Dx | 25 | 220 | 82.8 | 114.7 | -0.43 | 0.3 | 0.007 |
| MAGNESIUM, S/P | mg/dL | Days 1-3 Post-Dx | 17 | 161 | 2.45 | 1.93 | 1.24 | 0.26 | 0.008 |
| CARBOXYHEMOGLOBIN, ARTERIAL | % | Days 1-3 Post-Dx | 22 | 160 | 0.536 | 0.864 | -0.638 | 0.29 | 0.009 |
| TROPONIN T, 5TH GEN, P | ng/L | Days 1-3 Post-Dx | 12 | 23 | 130.1 | 741.4 | -0.622 | 0.18 | 0.013 |
| OXYGEN SATURATION (%) IN ARTERIAL BLOOD | % | Days 1-3 Post-Dx | 15 | 102 | 94.8 | 96 | -0.663 | 0.27 | 0.023 |
| LACTATE, P | mmol/L | Days 1-3 Post-Dx | 14 | 79 | 1.51 | 2.49 | -0.495 | 0.28 | 0.042 |
| SODIUM, P | mmol/L | Days 1-3 Post-Dx | 46 | 438 | 136.6 | 138.1 | -0.363 | 0.39 | 0.049 |
| RBC (RED BLOOD CELL) COUNT | x10(12)/L | Days 4-6 Post-Dx | 89 | 544 | 4.07 | 3.44 | 0.883 | 0.26 | 1.6397E-10 |
| HEMATOCRIT, B | % | Days 4-6 Post-Dx | 89 | 543 | 35.9 | 31.5 | 0.751 | 0.3 | 4.315E-08 |
| HEMOGLOBIN, B | g/dL | Days 4-6 Post-Dx | 89 | 518 | 11.7 | 10.2 | 0.743 | 0.3 | 1.0849E-07 |
| RED CELL DISTRIBUTION WIDTH CV | % | Days 4-6 Post-Dx | 75 | 444 | 14.1 | 15.6 | -0.583 | 0.31 | 3.5305E-06 |
| MONOCYTES ABSOLUTE | x10(9)/L | Days 4-6 Post-Dx | 73 | 376 | 0.445 | 0.711 | -0.594 | 0.31 | 1.2406E-05 |
| MEAN CORPUSCULAR VOLUME | fL | Days 4-6 Post-Dx | 89 | 544 | 88.7 | 92.6 | -0.545 | 0.34 | 5.9047E-05 |
| MAGNESIUM, S/P | mg/dL | Days 4-6 Post-Dx | 23 | 131 | 2.34 | 1.93 | 1.177 | 0.23 | 0.0005 |
| ALKALINE PHOSPHATASE, S | U/L | Days 4-6 Post-Dx | 35 | 229 | 86.9 | 153 | -0.398 | 0.29 | 0.001 |
| EOSINOPHILS ABSOLUTE | x10(9)/L | Days 4-6 Post-Dx | 50 | 299 | 0.0917 | 0.205 | -0.512 | 0.33 | 0.001 |
| CALCIUM, TOTAL, S | mg/dL | Days 4-6 Post-Dx | 49 | 352 | 8.33 | 8.7 | -0.589 | 0.34 | 0.003 |
| MAGNESIUM, PLASMA | mg/dL | Days 4-6 Post-Dx | 16 | 114 | 2.19 | 1.96 | 0.859 | 0.24 | 0.005 |
| RABG CALCULATED O2 HEMOGLOBIN | % | Days 4-6 Post-Dx | 9 | 20 | 92.8 | 95.8 | -1.888 | 0.11 | 0.007 |
| CALCIUM, TOTAL, P | mg/dL | Days 4-6 Post-Dx | 43 | 235 | 8.54 | 8.9 | -0.548 | 0.34 | 0.007 |
| BILIRUBIN, TOTAL, P | mg/dL | Days 4-6 Post-Dx | 17 | 118 | 0.429 | 1.36 | -0.508 | 0.26 | 0.01 |
| CARBOXYHEMOGLOBIN, ARTERIAL | % | Days 4-6 Post-Dx | 20 | 112 | 0.603 | 1.01 | -0.645 | 0.3 | 0.019 |
| OXYGEN SATURATION (%) IN ARTERIAL BLOOD | % | Days 4-6 Post-Dx | 14 | 69 | 93.9 | 95.7 | -0.867 | 0.27 | 0.029 |
| BICARBONATE, P | mmol/L | Days 4-6 Post-Dx | 43 | 228 | 23.8 | 25.5 | -0.449 | 0.37 | 0.036 |
| RBC (RED BLOOD CELL) COUNT | x10(12)/L | Days 7-9 Post-Dx | 68 | 388 | 4.02 | 3.44 | 0.776 | 0.29 | 2.4811E-06 |
| MAGNESIUM, S/P | mg/dL | Days 7-9 Post-Dx | 20 | 97 | 2.35 | 1.89 | 1.457 | 0.17 | 0.00007767 |
| ALBUMIN, P | g/dL | Days 7-9 Post-Dx | 21 | 122 | 2.94 | 3.35 | -0.666 | 0.28 | 0.007 |
| MONOCYTES ABSOLUTE | x10(9)/L | Days 7-9 Post-Dx | 52 | 288 | 0.472 | 0.686 | -0.462 | 0.37 | 0.014 |
| CALCIUM, TOTAL, S | mg/dL | Days 7-9 Post-Dx | 41 | 261 | 8.48 | 8.8 | -0.458 | 0.36 | 0.02 |
| ALBUMIN, S/P | g/dL | Days 7-9 Post-Dx | 39 | 197 | 3.36 | 3.63 | -0.441 | 0.36 | 0.021 |
| TACROLIMUS, B | ng/mL | Days 7-9 Post-Dx | 5 | 45 | 4.54 | 7.9 | -1.176 | 0.11 | 0.024 |
| CARBOXYHEMOGLOBIN, ARTERIAL | % | Days 7-9 Post-Dx | 18 | 75 | 0.662 | 1.04 | -0.593 | 0.3 | 0.037 |
| RBC (RED BLOOD CELL) COUNT | x10(12)/L | Days 10-12 Post-Dx | 54 | 293 | 3.88 | 3.4 | 0.655 | 0.32 | 0.0003 |
| MAGNESIUM, S/P | mg/dL | Days 10-12 Post-Dx | 17 | 66 | 2.24 | 1.86 | 1.288 | 0.2 | 0.001 |
| MEAN CORPUSCULAR VOLUME | fL | Days 10-12 Post-Dx | 54 | 293 | 89.2 | 92.9 | -0.514 | 0.34 | 0.001 |
| HEMATOCRIT, B | % | Days 10-12 Post-Dx | 54 | 288 | 34.4 | 31.2 | 0.528 | 0.35 | 0.003 |
| ALKALINE PHOSPHATASE, S | U/L | Days 10-12 Post-Dx | 31 | 153 | 102.2 | 182.6 | -0.38 | 0.31 | 0.008 |
| PLATELETS | x10(9)/L | Days 10-12 Post-Dx | 54 | 270 | 298.7 | 243.4 | 0.361 | 0.36 | 0.008 |
| MAGNESIUM, PLASMA | mg/dL | Days 10-12 Post-Dx | 9 | 62 | 2.2 | 1.87 | 1.283 | 0.17 | 0.009 |
| LACTATE DEHYDROGENASE, S | U/L | Days 10-12 Post-Dx | 10 | 62 | 434.1 | 323.4 | 0.383 | 0.2 | 0.015 |
| HEMOGLOBIN, B | g/dL | Days 10-12 Post-Dx | 54 | 293 | 11.1 | 10.2 | 0.415 | 0.38 | 0.02 |
| TRIGLYCERIDES, S/P | mg/dL | Days 10-12 Post-Dx | 7 | 12 | 541.9 | 245.2 | 1.706 | 0.13 | 0.042 |
| MAGNESIUM, S/P | mg/dL | Days 13-15 Post-Dx | 14 | 52 | 2.21 | 1.79 | 1.566 | 0.13 | 0.0004 |
| BUN, P | mg/dL | Days 13-15 Post-Dx | 12 | 96 | 49 | 24.1 | 1.454 | 0.23 | 0.013 |
| MEAN CORPUSCULAR VOLUME | fL | Days 13-15 Post-Dx | 34 | 266 | 90.3 | 93.9 | -0.511 | 0.34 | 0.016 |
| GLUPS GLUCOSE, P | mg/dL | Days 13-15 Post-Dx | 12 | 104 | 169.3 | 130.7 | 0.651 | 0.27 | 0.045 |
| BUN, P | mg/dL | Days 16-30 Post-Dx | 14 | 167 | 48.7 | 24.4 | 1.225 | 0.2 | 0.002 |
| MAGNESIUM, S/P | mg/dL | Days 16-30 Post-Dx | 10 | 93 | 2.21 | 1.83 | 1.223 | 0.19 | 0.009 |

**Table 3. Prevalence of thrombotic phenotypes in $COVID_{pos}$ patients with and without available longitudinal lab testing data.** For each clotting phenotype listed, a BERT-based neural network was used to extract diagnostic sentiment from individual EHR patient notes in which the phenotype (or a synonym thereof) was present. In bold: absolute number of patients with each phenotype; in parentheses: percentage of thrombotic patients in each cohort (i.e., column) with the given specific thrombotic phenotype; in brackets: percentage of all patients in each cohort with the given specific thrombotic phenotype.

| Clotting Phenotype | Cohort 1: $COVID_{pos}$ with longitudinal data | Cohort 2: $COVID_{pos}$ without longitudinal data | Cohort 3: Complete $COVID_{pos}$ cohort |
|---|---|---|---|
| *Deep vein thrombosis* | **25** (76%) [14%] | **2** (29%) [0.13%] | **27** (68%) [1.6%] |
| *Pulmonary embolism* | **8** (24%) [4.4%] | **3** (43%) [0.19%] | **11** (28%) [0.64%] |
| *Myocardial infarction* | **3** (9%) [1.7%] | **2** (29%) [0.13%] | **5** (13%) [0.29%] |
| *Venous thromboembolism* | **1** (3%) [0.55%] | **0** | **1** (2.5%) [0.06%] |
| *Thrombotic stroke* | **2** (6%) [1.1%] | **0** | **2** (5%) [0.12%] |
| *Cerebral venous thrombosis* | **0** | **0** | **0** |
| *Disseminated intravascular coagulation* | **2** (6%) [1.1%] | **0** | **2** (5%)[0.12%] |
| **Total Unique Patients with Clot** | **33** [18%] | **7** [0.84%] | **40** [2.3%] |
| **Total Patients** | **181** | **1548** | **1729** |

**Table 4. Enrichment of thrombotic phenotypes among $\text{COVID}_{pos}$ patients with longitudinal lab testing data.** Contingency table to calculate hypergeometric enrichment significance of thrombosis among patients with longitudinal lab testing data. The 181 patients with longitudinal testing data are those considered in this study, while the 1548 patients who did not have at least three results from one lab test over the defined 60-day window were excluded from this longitudinal analysis.

| | **Patient has longitudinal data** | **Patient does NOT have longitudinal data** | |
|---|---|---|---|
| **Thrombosis** | 33 | 7 | 40 |
| **No Thrombosis** | 148 | 1541 | 1689 |
| | 181 | 1548 | 1729 |

**Hypergeometric enrichment: p-value $< 1 \times 10^{-20}$**